\documentclass[useAMS,usenatbib,usegraphicx]{mn2e}

\usepackage{amssymb}
\usepackage{natbib}

\def \mstar {M$_{\star}$}
\def \lstar {L$_{*}$}
\def \msun {M$_{\odot}$}

\def \tnblue {$T_{N,blue}$}
\def \rvir {r$_{200}$}
\def \ks {Kolmogorov-Smirnov}

\title[Globular clusters and the origin of dEs in Virgo]{Globular cluster systems as tracers of environmental effects on Virgo early-type dwarfs}

\author[S\'anchez-Janssen \& Aguerri]{R. S\'anchez-Janssen\,$^{1}$\thanks{E-mail:
rsanchez@eso.org} and J.A.L. Aguerri\,$^{2,3}$\\
$^{1}$European Southern Observatory, Alonso de C\'ordova 3107, Vitacura, Santiago, Chile\\
$^{2}$Instituto de Astrof\' isica de Canarias, Calle V\'ia L\'actea s/n, E-38200 La Laguna, Tenerife, Spain\\
$^{3}$Departamento de Astrof\' isica, Universidad de La Laguna, E-38205 La Laguna, Tenerife, Spain}

\begin{document}

\date{2012 May 13}

\pagerange{\pageref{firstpage}--\pageref{lastpage}} \pubyear{2012}

\maketitle

\label{firstpage}

\begin{abstract}
Early-type dwarfs (dEs) are by far the most abundant galaxy population in nearby clusters. Whether these objects are primordial, or the recent end-products of the different physical mechanisms that can transform galaxies once they enter these high-density environments, is still a matter of debate. Here we present a novel approach to test these  scenarios by comparing the properties of the globular cluster systems (GCSs) of Virgo dEs and their potential progenitors with simple predictions from gravitational and hydrodynamical interaction models. We show that low-mass (M$_{\star} \lesssim 2\times10^{8}$ \msun) dEs have GCSs consistent with being the descendants of gas-stripped late-type dwarfs. On the other hand, higher mass dEs have properties --including the high mass specific frequencies of their GCSs and their concentrated spatial distribution within Virgo-- incompatible with a recent, environmentally-driven evolution. They mostly comprise nucleated systems, but also dEs with recent star formation and/or disc features. Bright, nucleated dEs appear to be a population that has long resided within the cluster potential well, but have surprisingly managed to retain very rich and spatially extended GCSs -- possibly an indication of high total masses. Our analysis does not favour violent evolutionary mechanisms that result in significant stellar mass losses, but more gentle processes involving gas removal by a combination of internal and external factors, and highlights the relevant role of initial conditions.
Additionally, we briefly comment on the origin of luminous cluster S0s.
\end{abstract}


\section{Introduction}
Early-type cluster dwarfs\,\footnote{Following \citet{Boselli2008}, by this definition we refer to any low-mass, quiescent cluster galaxy, thus including dwarf ellipticals, spheroidals and lenticulars. Throughout this paper they will all be termed dEs for simplicity -- but see \citet{Kormendy2009} for a  thorough discussion of the similarities and differences between these specimens.}  constitute a longstanding puzzle. Being the most abundant galaxy population in high-density environments \citep{Binggeli1985}, it is not yet fully understood whether they are old systems whose current properties were determined at the early stages of cluster assembly, or if they have a late origin as the end-products of the physical mechanisms that operate in virialised clusters. The scenario is indeed complex, as it is well known that dEs are not simple objects with uniform properties, but rather encompass a variety of systems with different stellar structure and age (e.g., \citealt{Ferrarese2006,Boselli2008,Lisker2009b} and references therein). It is therefore most likely that several formation mechanisms need to be invoked to explain the origin of these objects.

These mechanisms can be broadly divided into internal and external ones.
Among the first class, strong feedback effects due to supernovae and intense star formation is arguably the best candidate \citep{Dekel1986}, as it can lead to the removal of low angular momentum material --be it dark matter (DM) or the interstellar medium (ISM)--, potentially resulting in a thick system that is not DM-dominated in the central regions but can still be rotationally supported \citep{Governato2010}.
The distinct location of dEs on the Fundamental Plane appears to be consistent with this scenario  \citep{deRijcke2005} -- although environmental effects can not be ruled out \citep{Aguerri2009a}.

The second class includes a plethora of physical mechanisms that can shape galaxies in clusters (see \citealt{Boselli2006} for an excellent review). 
These external processes can be again subdivided in two groups: hydrodynamical interactions between the hot intracluster medium and the galaxy's ISM --e.g., ram-pressure stripping \citep{GunnGott1972} and starvation \citep{Larson1980}--, and gravitational interactions with the cluster potential and between galaxies themselves --tidal stripping \citep{Merritt1984} and tidal shocking \citep{Moore1996}.
In reality, all these mechanisms contribute to some extent to the evolution of cluster dwarfs, in a sort of tidal stirring process like the one proposed by \citet{Mayer2001a}.

Environmental effects are supported by a significant number of observational evidences, including the fact that a non-negligible fraction of dEs appear to be rotationally supported \citep{Pedraz2002,Geha2003,vanZee2004,Toloba2011}; the presence of disc features (spiral arms and/or bars) in a subpopulation of cluster dEs \citep{Barazza2002,Graham2003,Lisker2006,Mendez-Abreu2010,Janz2012}; the existence of both a bulge and disc components in the light profiles of some early-type dwarfs \citep{Aguerri2005,Kormendy2012}; or the strong transition towards older ages \citep{Haines2006,Smith2009} and redder colours \citep{rsj2008} of dwarfs within virialised cluster regions. 
Furthermore, the signatures of a late infall from cluster dwarfs dynamics \citep{Drinkwater2001,Conselice2001} and the realisation that the low-mass end of the red sequence progressively builds up at lower redshifts \citep{DeLucia2007} strongly suggest that this transformation is a relatively recent phenomenon that took place over the last $\sim$\,6\,Gyr.
Unfortunately, all these mechanisms can be efficient in multiple, sometimes coincidental cluster regions, and their timescales are also different. To observationally discriminate the predominant process(es) becomes a difficult task, requiring the use of many observables that can help break these degeneracies.

Globular clusters (GCs) are ancient systems found in almost all galaxy types along the Hubble sequence, irrespective of their luminosity or gas content \citep[BS06 hereafter]{Brodie2006}. Thus, the ubiquity of these stellar clusters have made them powerful tools for the study of star formation and mass assembly processes in nearby galaxies, as they hold a fossil record of their host galaxy evolutionary history. 
This fact has prompted the use of globular cluster systems (GCSs) as sensitive tracers of formation mechanisms and, to a lesser extent, environmental effects on different cluster galaxy populations \citep[e.g.,][]{AragonSalamanca2006,Miller2007,Coenda2009}. 
\citet{Peng2008} provide the most comprehensive analysis to date of the GCSs in Virgo dE galaxies, including an investigation of the roles of stellar mass and environment. They show that dEs in the inner cluster regions exhibit higher GC specific frequencies than those in the outskirts, but are indistinctive in terms of their structural properties or their stellar populations colours -- a behaviour they explain within a framework of an early origin for dEs and a biased GC formation towards dense environments.

In this contribution we extend their approach by quantitatively comparing the properties of the globular cluster systems of Virgo dEs and their potential progenitors with simple predictions from the gravitational and hydrodynamical interaction models that are traditionally thought to shape early-type cluster dwarfs. 
The Paper is organised as follows. In Section\,2 we present a brief overview of the data samples compiled from the literature that are used during the subsequent analysis. Section\,3 presents predictions of the joint evolution of galaxies and their GCSs from the most relevant interaction models in clusters. Sections\,4 is devoted to the constraints derived from observations, which are further discussed in Section\,5, together with future prospects that we consider can significantly improve our knowledge of dwarf galaxy formation and evolution.

\section{Data compilation}
In order to carry out this study, we have compiled data for the GCSs of cluster and field galaxies from several sources in the literature.
Most observational data come from the ACSVCS \citep{Cote2004}, an unprecedented HST survey of 100 early-type galaxies in the Virgo cluster spanning a wide magnitude range ($-23 < M_{V} < -15.5$). Taking advantage of the HST's unique resolution, they have carried out a thorough analysis of the GCSs in their sample, including studies of GC colour distributions and specific frequencies \citep{Peng2006,Peng2008}. 
Furthermore, \citet{Mei2007} derived surface brightness fluctuations distances for the majority of  the sample, allowing for a projection-free investigation of environmental effects as a function of position within Virgo.
Stellar masses, derived from colour fits to model single stellar populations, are also used as presented in \citet[P08 hereafter]{Peng2008}, and galaxies having \mstar\ $\lesssim 5\times10^{9}$ \msun\ are considered to be dwarfs.

In order to extend the analysis further into the low-mass regime, we have complemented the previous dataset with the faint ($-18 < M_{V} < -12.5$) Virgo dEs from \citet[ML07 hereafter]{Miller2007}. A mass-to-light ratio in the $V$-band of \mstar/L$_{V}=2$ is used when computing their stellar masses, consistent with those of the ACSVCS.
 
The properties of the GCSs of potential progenitors come from several sources. At the bright end we use the recent compilation of \citet[S08 hereafter]{Spitler2008}, that includes data for several M$_{\star} > 4\times10^{10}$ \msun\ elliptical and disc galaxies. 
This dataset is particularly well suited for our analysis, as it only contains galaxies having total GC number uncertainties lower than 40\%, a large spatial coverage (more than 50\% of the GCS radial extent), available information on metallicity subpopulations proportions,  and self-consistently derived stellar masses. We further complement this dataset with the late-type galaxy compilation of \citet{Chandar2004}, using the morphological type-dependent \mstar/L$_{V}$ ratios derived by S08. Given that the majority of galaxies in the sample lack information regarding subpopulation fractions, we assume they all consist of metal-poor GCs. These values are therefore much more uncertain than the S08 ones (and, probably, upper limits), but we nonetheless include them in our analysis for comparison purposes.

At the faint end we use data from the recent work of \citet[G10 hereafter]{Georgiev2010}, consisting of low-mass (M$_{\star} < 3\times10^{9}$ \msun) galaxies in the Local Universe ($D \lesssim 10$ Mpc). From this work we only select late-type (T $\ge$ 5) dwarfs, and use stellar masses as tabulated by the authors. Additionally, we also include the seven Virgo dIrrs studied by \citet[S04]{Seth2004}, for which \mstar/L$_{V} = 1$ is assumed -- consistent with both the G10 and P08 values.

We note that these samples comprise the most homogeneous and robust compilations of GCSs for their respective galaxy types. This is especially true in the case of dwarf galaxies, where all four studies benefit from the unrivaled resolution and sensitivity of the HST.

\section{Environmental Effects on Galaxies and Their Globular Cluster Systems: Model Predictions}
Globular clusters, with their predominantly old ages\,\footnote{The simulations of \citet{Beasley2002} predict mean ages of 12 and 9 Gyr for the metal-poor and metal-rich subpopulations, respectively -- though the latter are characterised by a much broader age distribution. This is consistent with observations of both Galactic and extragalactic GCs (see BS06 and references therein). In the particular case of the Virgo cluster, old GC ages ($>$10 Gyr) have been inferred from spectroscopic observations for galaxies spanning two orders of magnitude in stellar mass, from the giants M87 \citep{Cohen1998} and M49 \citep{Puzia1999}, to the dE VCC\,1087 \citep{Beasley2006}.}, have been privileged witnesses of a significant fraction of their host galaxy evolutionary history. This, together with their intrinsic compact sizes and extended galactocentric distributions, makes them ideal probes of interaction processes in galaxy clusters. For instance, they are in principle insensitive to all the hydrodynamical mechanisms that can alter the properties of a galaxy's ISM once it enters the cluster environment. On the other hand, gravitational interactions can modify the stellar content of galaxies, and this should be reflected on their GCSs as well. Here we make predictions of the joint evolution of galaxies and their GCSs using basic theoretical considerations of the most relevant interaction models in clusters, i.e., gravitational harassment and gas stripping.
We focus on relatively recent interactions ($\lesssim$\,6 Gyr), and implicitly assume that none of these processes lead to GC formation.

\subsection{Gravitational Harassment}
\label{sect:harass}
One of  the most popular evolutionary paths for cluster dEs is their transformation from late-type galaxies through multiple, fast gravitational encounters within the cluster potential, i.e., harassment \citep{Moore1996}. 
This tidally-induced mechanism is supposed to be rather efficient at stripping gas and stars from both relatively massive, low surface brightness galaxies (LSBs; \citealt{Moore1999}) and lower-mass dwarf irregular (dIrr) and Magellanic-types \citep{Mastropietro2005}. The resulting remnants closely resemble observed cluster dEs (e.g., \citealt{Aguerri2009a}).

Gravitational interactions in clusters like Virgo are sufficiently rapid that they can be treated under the impulse approximation \citep{Spitzer1958}. These fast encounters result in a net increase of the kinetic energy per unit mass of the target's particles in the form (e.g., \citealt{Aguilar1985}):

\begin{equation}
\label{eq:impulse}
(\Delta E/m) = G^{2}\,M_{P}^{2}\,v^{-2}\,b^{-4}\,r^{2}\,f(P,A), 
\end{equation}

\noindent
where $G$ is the gravitational constant, $M_{P}$ is the perturber's mass, $v$ is the encounter velocity, $b$ is the impact parameter, and $r$ is the distance of the affected particle to the target's centre. This energy change is further modulated by the $f(P,A)$ term, that includes a correction for both the extended nature of the perturber, $f(P)$, and for the movement of the stars within the target -- the adiabatic correction. The latter can take the form $f(A) = (1+x^{2})^{-\gamma}$, where $x=\omega\tau$ is the product of the orbital frequency of particles in the target ($\omega$) and the effective duration of the encounter ($\tau$), and $1.5 < \gamma < 3$ depending on the duration of the shock with respect to the target's half-mass dynamical time \citep{Gnedin1999a}. 
Equation\,\ref{eq:impulse} implies that the energy gain is maximal for particles at large galactocentric distances with small orbital frequencies, while it is strongly diminished if they are concentrated or if tidal forces act on a time scale much larger than their orbital period. The larger the kinetic energy acquired, the less bound particles are. As a result, the strong dependence on galactocentric distance ($\propto$\,$r^{2}$) implies that in the impulsive regime mass loss preferentially occurs from the outside-in. 

The efficiency of mass stripping due to harassment depends on the collisional frequency, and on the strength of individual collisions and of the cluster tidal field. \citet{Moore1998}  show that the competing effects of these terms result in harassment being a rather effective mechanism in a wide range of clustercentric radii (see also \citealt{Gnedin2003}), but more so closer to the cluster centre, where both the cluster tidal field and the collision rate are expected to be maximal.
Taking all this into account, we can make the following predictions for the evolution of GCs and the stellar body in harassed galaxies:
\begin{itemize}
\item[i)] The detailed stripping history of both components will be determined by their phase-space (spatial and orbital) distribution. If GCs have a more extended spatial distribution and lower rotational support than stars, they should be stripped more efficiently (e.g., \citealt{Bekki2003}) -- and viceversa.
\item[ii)] The effects of mass stripping should present a clustercentric dependence, in the sense that objects that spend longer times close to the centre are affected the most (e.g., \citealt{Mastropietro2005,Smith2010}).
\end{itemize}

\subsection{Gas Stripping}
\label{sect:gas}
The alternative and perhaps most obvious scenario for the origin of dEs is gas removal from a population of infalling, field dIrrs. Gas sweeping can occur through several processes that act on different timescales, including ram-pressure \citep{GunnGott1972} and viscous \citep{Nulsen1982} stripping, starvation \citep{Larson1980} and thermal evaporation \citep{Cowie1977}. They all can halt star formation and --especially ram-pressure-- create low-mass, passive galaxies structurally and chemically compatible with those observed in Virgo \citep{Boselli2008}.

Whatever the specific mechanism, it is important to note that their effect on the already-existing stellar body is essentially negligible \citep{Quilis2000}, and the same is true for the GCS.
Therefore, if gas-stripping mechanisms play a predominant role we expect to see no difference whatsoever between the GCSs of potential progenitors and remnants. 
One important caveat is that this is not exactly true for their stellar mass content. Indeed, while star formation in field dIrrs has continuously increased \mstar, in the gas-stripping plus star formation shutoff scenario the current masses of dEs should essentially reflect those of their progenitors upon cluster arrival. This stellar mass difference will generally depend on the detailed star formation history of the progenitor, and on the efficiency and timescale of the stripping process -- resulting in a 'progenitor bias' that has to be taken into account.

\section{Constraints from Observations}

\subsection{GC Mass Specific Frequencies}
One of  the most  constraining observables provided by GCSs is their specific frequency, i.e., the number of GCs normalised by galaxy luminosity or stellar mass. If GCs indeed form with a universal efficiency \citep{McLaughlin1999b}, variations of specific frequency  among galaxy types might reflect a different evolutionary history. For example, if  dEs are the descendants of transformed galaxies, the properties of both GCSs must be somehow related -- and if several evolutionary scenarios lead to the formation of these objects we expect their distinct imprint to be recognisable on their GCSs.

Here we explore the connection between dEs and their potential progenitors by means of the GC mass specific frequency, $T_{N} = N_{gc}/(\mbox{\mstar}/10^{9}\,\mbox{\msun})$. \citet{Zepf1993} modified the classic $S_{N}$ definition of \citet{Harris1981} --replacing the galaxy luminosity by its stellar mass--, thus allowing for a more straightforward comparison between galaxies with different \mstar/L ratios. 
Moreover, this variation turns out to be a better discriminator between interaction types than $S_{N}$. 
From the parameters involved in these definitions, it is clear that both $T_{N}$ and $S_{N}$ are expected to vary during a gravitational encounter involving any kind of stellar mass loss (stars or GCs). Neither the stellar mass of a galaxy nor its number of GCs are modified during a gas stripping event -- thus leaving $T_{N}$ unchanged. But the lack of gas reservoirs for future star formation unavoidable leads to luminosity fading  --and $S_{N}$ varying-- due to stellar populations aging. 
In other words, $S_{N}$ variations are expected due to both gravitational and hydrodynamical processes, while $T_{N}$ is only sensitive to the former.

During this discussion we further focus on the mass specific frequency of metal-poor (blue) GCs. This subpopulation, unlike the metal-rich (red) one, is unambiguously old and appears to be ubiquitous across galaxies of all  masses and morphological types -- therefore allowing for a direct comparison between dEs and their potential progenitors. Furthermore, they are traditionally associated to the halo of L $\gtrsim$ \lstar~galaxies due to their extended spatial distribution and kinematics (BS06), making them excellent tracers of gravitational interactions (cf. Sect.\,\ref{sect:harass}).

\begin{figure*}
\begin{center}
\includegraphics[width=0.9\textwidth]{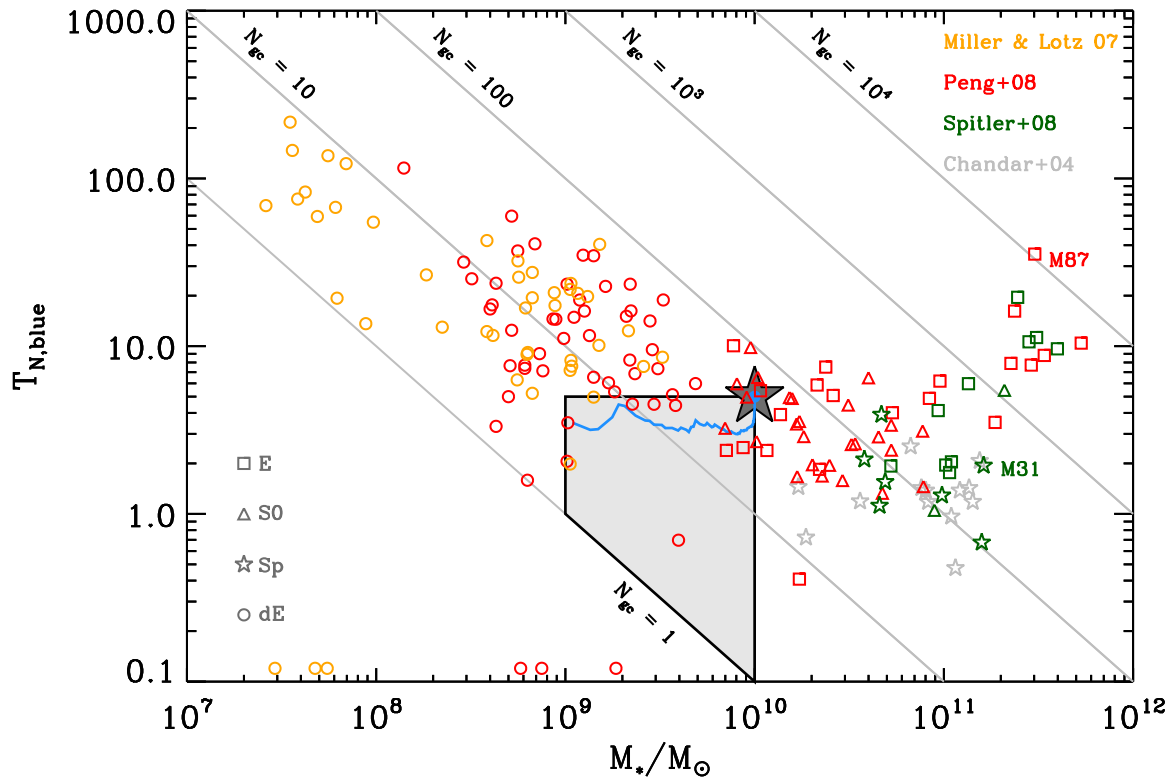}
\caption{Mass specific frequency of blue (metal-poor) GCs as a function of stellar mass for E, S0, Sp and dE galaxies from the literature. Galaxies having no GCs have been assigned a $T_{N,blue} =0.12$, and diagonal lines indicate the loci of galaxies having from one to 10$^{4}$ blue GCs. The solid star shows the initial $T_{N,blue}$ of our fiducial LSB model, while the shaded region indicates its potential location according to our scaling of the harassment model of \citet{Moore1999}. The blue curve indicates its expected evolution if mass stripping occurs from the outside-in (see text for details).
The only early-type dwarfs compatible with being harassment remnants are those showing low $T_{N,blue}$ values. LSB galaxies would need to lose more disc stars than GCs in order to match the high specific frequencies observed in most dEs.}
\label{fig:harass_lsb}
\end{center}
\end{figure*}

In Fig.\,\ref{fig:harass_lsb} we present the distribution of \tnblue\ for metal-poor clusters as a function of \mstar\ for Es, S0s, Sps and dEs in our compilation. Diagonal lines correspond to the location of objects having from one to 10$^{4}$ blue GCs. 
The well-known non-monotonic relation between \tnblue\ and galaxy stellar mass is clearly depicted in this plot:  intermediate-luminosity galaxies exhibit rather uniform, low values (note the logarithmic scale), while there is a clear trend for both low-mass and high-mass galaxies to have increasing specific frequencies. This already hints at the possibility that a transition from the intermediate-mass to the low-mass regime most likely requires an increase in \tnblue. 
We develop on this scenario by providing a more rigorous comparison between the GCSs of cluster dEs and their potential progenitors.
It is important to note that evolutionary paths that decrease (increase) the number of GCs of a galaxy necessarily have shallower (steeper) slopes than the constant $N_{gc}$ diagonal lines in Fig.\,\ref{fig:harass_lsb}. 

\subsubsection{LSBs vs dEs}
\label{sect:tn_lsb}
We start by comparing the GC mass specific frequency of LSBs and dEs with a simple scaling of the harassment models of \citet[][M99 hereafter]{Moore1999}. These authors follow the dynamical evolution of a \mstar\ $\sim 10^{10}$ \msun\ LSB within a massive galaxy cluster. Given that GCSs were not included in the simulations of M99, we estimate their evolution as follows.

The GCSs of luminous disc galaxies are characterised by having rather homogeneous properties. Their specific frequency appears to be almost universal (\citealt{Goudfrooij2003}; \citealt{Chandar2004}; see also Fig.\,\ref{fig:harass_lsb}), and they have a more extended spatial distribution than disc stars --especially the metal-poor population  (see BS06). Interestingly, these features appear to be \emph{independent} of  galaxy surface brightness, as indicated by the analysis of GCSs in five galaxies spanning a wide range of $\mu_{0}$ \citep[][V08 hereafter]{Villegas2008}.
We stick to these results and assume that the properties of the GCSs of intermediate-luminosity LSBs are similar to those of high surface brightness (HSBs) disc galaxies like M31 or the Milky Way.\,\footnote{We stress that our assumption only concerns the similarity of the GCS and not of the galaxy as a whole. M99 show that HSBs barely lose stellar mass through harassment and therefore can not evolve into dEs -- but they might do into S0s (see Appendix\,\ref{appendix}).}
Fig.\,\ref{fig:harass_lsb} illustrates the narrow range of frequencies characteristic of disc galaxies, all having $1 \lesssim $ \tnblue\ $ \lesssim 4$ (see also \citealt{Rhode2007}). 
In order to match the general trend in Fig.\,\ref{fig:harass_lsb} we assign our $10^{10}$ \msun\ model LSB a fiducial mass specific frequency \tnblue\ $=5$ (large solid star) -- or, equivalently, an spatially extended population of $N_{gc,blue} = 50$. 
This is a generous \emph{upper} limit to the metal-poor GC specific frequencies observed in disc galaxies.
M31 and M87, with \tnblue\ $\approx$ 2 and 35, respectively, are labeled for reference.

We noted in Section\,\ref{sect:harass} that the detailed stripping histories of stars and GCs will critically depend on their spatial and kinematical properties, which are poorly constrained for disc galaxies outside of the Local Group. Fortunately, we can combine the expectations from the impulse approximation and results from numerical simulations to provide an order-of-magnitude estimate of their evolution.  
Observations indicate that the metal-poor GC population in \lstar\ disc galaxies has a more extended spatial distribution than disc stars. Additionally, they most likely are less extended than particles of the DM halo. These two constraints to the metal-poor GCS spatial distribution translate into the expectation that GCs should be more easily stripped than stars, but less than DM. 
Moreover, this probably holds even at a fixed radius, as kinematical studies of the best studied disc galaxies --M31 and the MW, see BS06-- indicate that GCs move on more (less) eccentric orbits with less (more) rotational support than disc stars (DM). 
Thus, we generically predict that the fraction of stripped GCs will always be intermediate to that of the DM halo and the bulk of the stellar disc ($f_{dm} \ge f_{gc} \ge f_{disc}$). 
In order to provide a more quantitative  comparison with observations, we use the results from the harassment simulations of M99 (their table\,1), where it is shown that up to 90\% of disc stars and 95\% of DM mass can be stripped -- i.e., $0 < f_{disc} < 0.9$ and $0 < f_{dm} < 0.95$.

The solid star in Fig.\,\ref{fig:harass_lsb} shows the initial specific frequency of the model LSB, while the shaded region indicates its potential location according to our evolutionary predictions. 
The horizontal upper limit corresponds to the case where the LSB loses the same fractions of GCs and disc mass, hence \tnblue\ remains constant. 
The rightmost and leftmost vertical boundaries indicate the extreme cases where the stellar mass losses are 0\% and 90\% of the original value, respectively -- but the fraction of stripped GCs is always larger and therefore \tnblue\ decreases.
The lower diagonal boundary is the limiting case where the galaxy is left with  one GC only, while the stellar mass loss ranges between the two chosen extremes.
Interestingly, the low-mass, high-\tnblue\ region of the diagram is not accessible by this type of interaction, implying that such dEs are difficult to understand within the harassment scenario. The only way an LSB remnant can increase its \tnblue\ is by preferentially losing more disc stars than GCs, an unlikely evolution according to the impulse approximation. 
Alternatively, LSBs must have very different GCS properties (abnormally high frequencies or a  very concentrated spatial distribution) than the bulk of the disc galaxy population -- but this is not supported by observations.

The blue curve in Fig.\,\ref{fig:harass_lsb} provides a more specific prediction of the evolution of harassed LSB galaxies. We examine the limiting case where stripping occurs strictly from the outside-in, i.e., without taking the orbital properties of stars and GCs into account. The model consists of a scaled version of the Milky Way. We assume that the metal-poor clusters have the same spatial distribution as in the MW -- a standard practice in studies of extragalactic GCSs in disc galaxies (e.g., \citealt{Goudfrooij2003}; \citealt{Chandar2004}; V08). We use the Galactic GCS data compiled by \citet{Bica2006}, and select metal-poor clusters as those with [Fe/H] $\le -1$. 
The LSB model has a \mstar\ $= 10^{10}$\,\msun\ disc with the same exponential scale length as the MW ($h = 2.5$ kpc)  -- and therefore, a factor 10 lower central surface mass density.  
At any given radius we estimate the fractions of enclosed GCs and disc mass, and the (properly normalised) ratio provides the evolution of the mass specific frequency. 

As expected, \tnblue\ rapidly decreases at first due to the outermost GCs being stripped without significant stellar mass loss taking place. Stripping then continues with more or less equal efficiency for both components, resulting in a low-mass, low-\tnblue\ remnant.
This is admittedly a simplistic approximation, as we assume that the spatial distribution of stars and GCs are not altered during the evolution, but nonetheless shows that it is extremely difficult to increase \tnblue\ if stripping occurs from the outside-in. It is of course possible that, as mass loss proceeds, GCs in the inner regions start to become more resilient to stripping than stars. What Fig.\,\ref{fig:harass_lsb} shows is that at this point the number of surviving GCs is already too low ($N_{gc,blue} \lesssim 10$) to match the rich GCSs observed in most massive dEs. And, given that metal-poor GCSs tend to be less rotationally supported than disc stars (BS06), we can further expect this \tnblue\ evolution to be an \emph{upper} limit to the actual process.

We finally note that, most likely, the GCSs of Virgo dEs  have also been affected by the cluster tidal field during their recent orbital evolution,  resulting in some level of stripping and/or enhanced orbital decay \citep{Oh2000}. If their GCSs were richer in the past, the mass specific frequencies currently observed can be thought of as lower limits -- thus making the disagreement  with disc galaxies even stronger.

\subsubsection{dE subpopulations, dIrrs, and the importance of environment}

\begin{figure*}
\begin{center}
\includegraphics[width=0.9\textwidth]{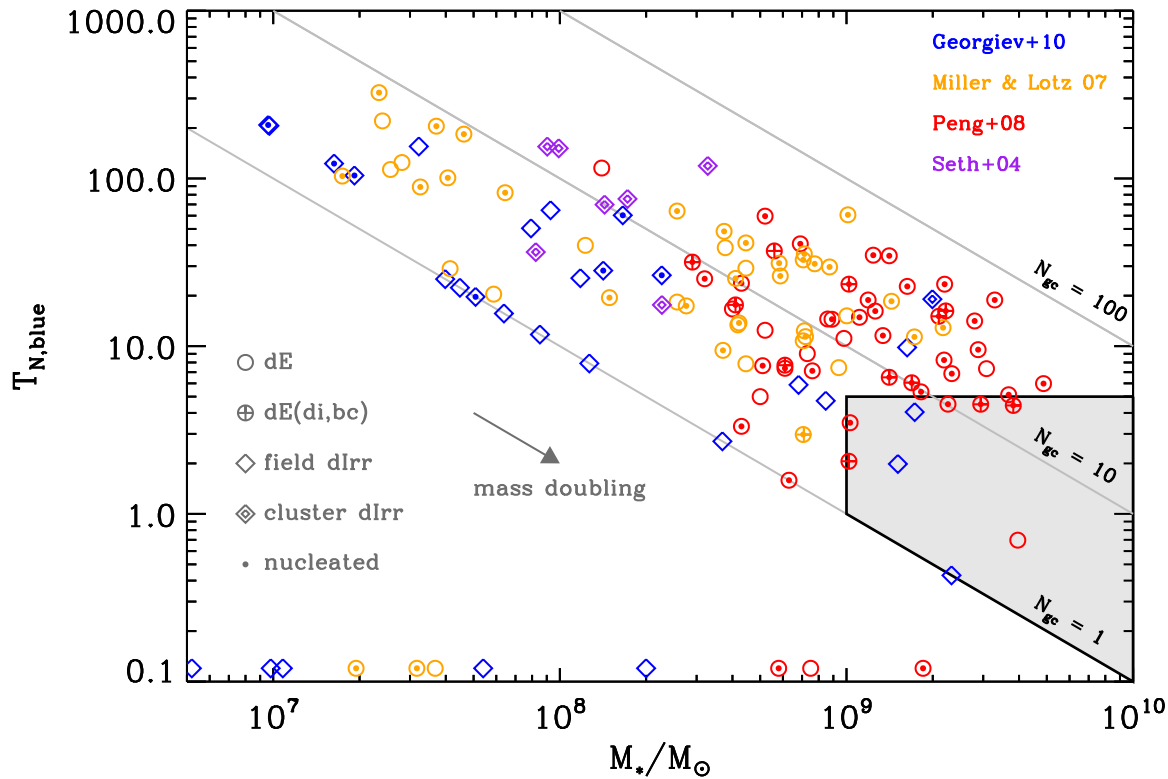}
\caption{Mass specific frequency of metal-poor GCs for cluster and field dIrrs, and Virgo dEs -- including both nucleated and non-nucleated objects, as well as systems with disc features and central star formation. Galaxies having no blue GCs have been assigned a \tnblue\ $=0.12$, while diagonal lines correspond to the loci of systems with 1, 10 and 100 blue GCs. The shaded region indicates, as in Fig.\,\ref{fig:harass_lsb}, the potential location of harassed disc-like galaxies. Late-type \emph{field} dwarfs preferentially have lower specific frequencies than both late- and early-type \emph{cluster} dwarfs -- a feature that is more prominent at the high-mass end and holds even for nucleated dIrrs. The arrow indicates the effect of \mstar\ doubling had the dE progenitors not had their star formation quenched (see text for details).} 
\label{fig:dwarfs_gcs}
\end{center}
\end{figure*}

Recent observational work indicate that, on average, early-type dwarfs have much richer GCSs than late-types (e.g., ML07, G10). This is illustrated in Fig.\,\ref{fig:dwarfs_gcs}, a zoomed-in version of Fig.\,\ref{fig:harass_lsb} where we show the mass specific frequency of metal-poor GCs for Virgo dEs (ML07, P08) and dIrrs (S04), and field dIrrs from G10.
There are several features in this Figure worth discussing.

First, the richest GCSs among dEs are always found at the high-mass end: essentially all galaxies having more than ten blue GCs are more massive than \mstar\ $\approx 2\times10^{8}$. This results in a significant GC abundance difference with respect to \emph{field} dIrrs, where only 2/33 reach that number. 
Remarkably, dIrrs in the Virgo cluster appear to be richer than their counterparts in low-density environments, and they nicely match the high specific frequencies of dEs. 
In the narrow $5\times10^{7} <$ \mstar/\msun\ $< 5\times10^{8}$ mass range, field dIrrs have a median number of $N_{gc,blue} = 3$ metal-poor GCs, compared to $N_{gc,blue} = 13$ for Virgo dIrrs (see Fig.\,\ref{fig:dwarfs_gcs}). 
This difference was first uncovered by S04, and argues in favour of a fundamental difference between field and cluster late-type dwarfs in general. For instance, NGC\,1427A --the dIrr galaxy having the richest GCS in Fig.\,\ref{fig:dwarfs_gcs} from the G10 sample-- actually \emph{is} a member of the Fornax cluster.

We recall that a direct comparison between early- and late-type dwarfs is hampered by the different star formation histories (i.e., stellar mass growth) of these populations (cf. Sect.\,\ref{sect:gas}). To take this potential 'progenitor bias' into account, the arrow in Fig.\,\ref{fig:dwarfs_gcs} indicates the effect of \mstar\ doubling had the dE progenitors not had their star formation quenched. In this case galaxies simply would move along constant $N_{gc}$ paths, and this does not change the fact that most dEs contain way too many GCs compared to field dIrrs.

We have so far not distinguished between the different subpopulations in which Virgo dEs can be divided. As discussed by \citet{Lisker2007}, there are at least four subclasses of dEs with distinct morphological and clustering properties: dE(bc), with central star formation; dE(di) exhibit disc features;  dE(nN) have weak or no nuclei; and the classic nucleated dE(N). Put in this order, they can be naively thought of as forming a temporal sequence of cluster membership, from the most recent newcomers to the oldest cluster inhabitants. It is therefore pertinent to investigate whether these different properties also extend to their GCSs, as testament to their evolutionary histories.

In Fig.\,\ref{fig:dwarfs_gcs} we adopt the classification of \citet{Lisker2007} for each dE, but group dE(bc) and dE(di) together for the sake of statistics. Even though the ACSVCS imaging has demonstrated the presence of nuclei in galaxies that were thought not to have one from ground-based studies \citep{Cote2006}, here we maintain the VCC nuclear classification for homogeneity reasons -- ML07 did not re-classify their galaxies. We therefore stick to \citet{Lisker2007} and consider as nucleated only those dEs having a 'significant' nucleus. Furthermore, and in order to minimise the impact of stellar mass effects, we only focus on galaxies in their 'bright' dE subsample -- roughly corresponding to systems having \mstar\ $> 2\times10^8$ \msun.

\begin{table}
\caption{Statistical properties of the metal-poor GCSs in the different \mstar\ $>2\times10^{8}$ \msun\ dwarf types in Fig.\,\ref{fig:dwarfs_gcs}. Each non-diagonal element in the upper table indicates the confidence level, according to a \ks\ test, with which we can rule out the null hypothesis that the two \tnblue\ distributions are drawn from the same parent population. The diagonal elements show the fraction of galaxies of each type that have ten or more blue GCs. In the lower table we show the corresponding lower quartiles, medians and upper quartiles of the \tnblue\ distributions.}
\label{tab:tn}
\centering
\begin{tabular}{lrrrr}\hline 
$f(\ge$\,10\,GCs) & dIrr & dE(di,bc) & dE(nN) & dE(N) \\ \hline\hline 
dIrr           & \emph{12} & 78.8  & 97.5   & 99.8 \\  
dE(di,bc) & 78.8  & \emph{46} & 63.4   & 92.3 \\   
dE(nN)    & 97.5  & 63.4   &  \emph{50} & 80.3 \\  
dE(N)      & 99.8  & 92.3   & 80.3   &   \emph{78} \\ \hline\hline
$Q_{1}$(\tnblue) & 3 & 5 & 7 & 11 \\
$Q_{2}$(\tnblue) & 5 & 8 & 12 & 19 \\
$Q_{3}$(\tnblue) & 10 & 18 & 18 & 33\\ \hline
\end{tabular}
\end{table}

Table\,\ref{tab:tn} summarises the main properties of metal-poor GCSs in these three dE subclasses, plus the field dIrrs of G10 with stellar masses larger than the already mentioned value. 
The non-diagonal elements in the upper part of the table contain the results of \ks\ tests for the corresponding \tnblue\ distributions, while the diagonal elements show the fraction of each population having ten or more blue GCs. The lower part of Table\,\ref{tab:tn} provides a more quantitative description of the distributions through their lower quartiles, their medians and their upper quartiles.
The trend is unequivocal, and shows that the \emph{earlier} the galaxy type, the richer its GCS. 
dE(N) contain, by far, the most populous systems and they appear to be inconsistent with those of field dIrrs at the 3\,$\sigma$ level. The sole presence of bright nuclei in dE(N) would already argue in favour of this distinction, but it is important to note that this difference appears to hold for fainter nucleated dEs and dIrrs too (see Fig.\,\ref{fig:dwarfs_gcs}).
dE(nN) and dE(di,bc) come next, and display rather similar frequencies.\,\footnote{It is important to recall that the \ks\ test is specially sensitive to the median of the distributions, and hence the probability differences of these two subpopulations in Table\,\ref{tab:tn}.}
Table\,\ref{tab:tn} shows that dE(nN), as also found by ML07, systematically have poorer GCSs than their nucleated counterparts -- but their \tnblue\ distribution is still inconsistent with that of dIrrs at  $>$\,2\,$\sigma$ level.
The dE(di,bc) subclass is of particular relevance. They are characterised by the presence of spiral arms, bars and/or signs of recent star formation, and all these suggest they might be the most clear-cut examples of a recent environmental transformation. 
Indeed, Fig.\,\ref{fig:dwarfs_gcs} shows that a few of these systems marginally occupy the region where we expect the harassed remnants of more luminous disc-like galaxies to be found.
Surprisingly, though, the rather elevated upper quartile of their frequency distribution, $Q_{3}$(\tnblue) = 18, indicates that a non-negligible fraction of these dEs have significantly populous metal-poor GCSs. For instance, VCC\,856 --the paradigmatic example of a dE with spiral structure \citep{Jerjen2000}-- has \tnblue\ = 16, a factor four larger frequency than the Sp with the highest \tnblue\ in Fig.\,\ref{fig:harass_lsb}.

The paucity of metal-poor GCs in field late-type dwarfs represents a challenge for \emph{recent} environmentally-driven evolutionary scenarios. Fading of infalling dIrrs after a recent ($<$\,6 Gyr) gas stripping event does not change the number of GCs, and is thus incompatible with the rich systems we observe in \mstar\ $\gtrsim 2\times10^{8}$ \msun\ dEs -- but appears to be fully consistent at lower masses.
Could gravitational interactions then be responsible? We already pointed out that for harassment to increase the specific frequency of a given galaxy it is required that mass from the  stellar body is preferentially lost with respect to the GC population. This can happen if, for example, GCSs have a similar or a more concentrated spatial distribution than stars do. 
This is a priori possible for low-mass late-type galaxies, as G10 show that the GCSs in dIrrs are essentially concentrated within the galaxy stellar body ($1-2$\,r$_{e}$; see also \citealt{Mora2007}).
However, even if this leads to an increase of \tnblue, harassment does not provide a satisfactory solution. 
It is clear from Fig.\,\ref{fig:dwarfs_gcs} and Table\,\ref{tab:tn}  that the number of metal-poor GCs in field dIrrs seems to be too low to reproduce the GCSs of most massive dEs even without invoking any kind of mass loss. The richer GCSs found in cluster dIrrs alleviate this tension, but imply that these two populations of star-forming dwarfs most likely had very different initial star formation histories -- and that cluster dE progenitors are intrinsically different from field dwarfs \citep[e.g.,][]{Skillman1995}.

\subsection{GC Metallicity Distribution}
\begin{figure}
\begin{center}
\includegraphics[width=0.5\textwidth]{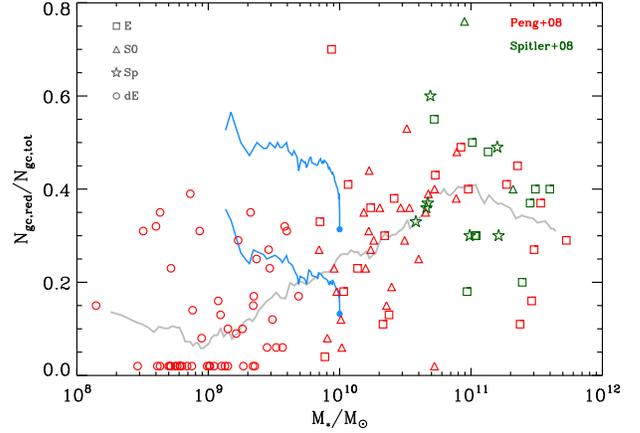}
\caption{Fraction of red (metal-rich) GCs as a function of host galaxy stellar mass. Galaxies consistent with having no red clusters have been slightly offset for displaying purposes. The gray curve is a running average, and shows that there is a trend for \mstar\ $\lesssim 10^{11}$ \msun\ galaxies to have lower metal-rich GC proportions. Blue curves indicate the expected evolution of a $10^{10}$ \msun\ galaxy if mass stripping occurs from the outside-in (see text for details). In both cases the galaxy hosts a GCS with subpopulations having an identical spatial distribution to that of the MW, and initial red fractions of $\approx$\,30\% and $\approx$\,15\%. Contrary to observations, the red GC fraction would tend to increase towards lower masses -- their less extended spatial distribution makes them more resilient to stripping than the blue subpopulation.}
\label{fig:harass_fred_evol}
\end{center}
\end{figure}

A second constraint can be derived from the GC metallicity distribution function of dEs and potential progenitors. It is well established that the GCSs of early-type galaxies present a bimodal metallicity (colour) distribution, with the fraction of metal-rich (red) clusters decreasing, and becoming almost negligible, towards fainter galaxy masses/luminosities \citep{Peng2006}. 
Disc-dominated \lstar~galaxies also contain a relatively important metal-rich population  that usually accounts for more than a third of their GCS.
This is shown in Fig.\,\ref{fig:harass_fred_evol}, where we plot the fraction of red (metal-rich) clusters as a function of \mstar\ for the two samples (P08 and S08) having accurate estimates of subpopulations proportions

The red GC fraction peaks at \mstar\ $\approx 10^{11}$ \msun, and decreases towards both higher and lower masses. The majority (22/37) of dEs with \mstar\ $\lesssim 2\times10^{9}$ \msun\ are consistent with having no red GCs at all.
The wide range of metal-rich fractions displayed by dEs might suggest that these galaxies followed alternative evolutionary paths, or their progenitors were intrinsically different.

\subsubsection{LSBs vs dEs}
Therefore, while massive galaxies contain a significant population of metal-rich GCs, these red clusters appear to be less abundant in early-type dwarfs.
In this context, for dEs to evolve from more massive systems it would require the latter to preferentially lose red GCs, so their relative fractions can be matched. 
This is in principle a difficult task as, at least in disc galaxies such as M31 and the Milky Way, the metal-poor population is much more extended than the metal-rich one (e.g., \citealt{Bica2006}) -- and LSBs appear to be consistent with this picture (V08). Given the strong galactocentric dependence of energy gain during an impulsive encounter, we would expect just the opposite trend --i.e., that the relative fraction of red clusters tends to increase. We again quantify this using the same MW scaling as in Sect.\,\ref{sect:tn_lsb}: at any given radius we estimate the fraction of enclosed disc mass, metal-rich ([Fe/H] $> -1$) and metal-poor ([Fe/H] $\le -1$) clusters. 

The two blue curves in Fig.\,\ref{fig:harass_fred_evol} show the evolution of the red GC fraction during this simple outside-in stripping model. The upper curve corresponds to an initial red GC proportion of $\approx$\,30\%, while for the lower one we randomly selected half of the original metal-rich subpopulation. It is clear that the more concentrated spatial distribution of red clusters leads to an enhanced relative fraction with respect to the reference value. 
If some dEs are the remnants of harassed LSBs, they have to be found among the high red GC fraction specimens.
The discrepancy might however be alleviated if we consider that LSB progenitors can have a low bulge mass fraction -- or be actually bulge-less. As metal-rich GCs are traditionally associated to the bulge component in disc galaxies \citep{Forbes2001}, it would be possible that these systems possess a negligible fraction of red GCs, as for example has been observed in M51 (\citealt{Chandar2004}; but again see V08 for some evidence of red GCs in LSBs).

\subsubsection{dIrrs vs dEs}
A comparison with the GCS metallicity distribution of low-mass, late-type galaxies is perhaps more puzzling. G10 found a significant fraction of dIrrs with an abundant --and sometimes even \emph{dominant}-- metal-rich subpopulation that could not be accounted for by contamination in their sample. 
Metal-rich GCs are not expected in low-mass systems from theoretical considerations on their formation process, and Fig.\,\ref{fig:harass_fred_evol} indicates that they are actually not observed in a significant fraction of Virgo dEs. G10 suggest that very young ages or internal galactic extinction could be responsible for this effect.
If metal-rich clusters turn out to be a significant subpopulation in dIrrs, it would be difficult to understand, in terms of environmental effects, how did these galaxies evolve into the metal-poor-dominated dEs we see in Virgo. Alternatively, the discrepancy between the metal-poor \tnblue\ values of both galaxy types (Fig.\,\ref{fig:dwarfs_gcs}) might be somewhat alleviated if reddening is responsible for this effect. We however note that spectroscopic observations indicate that at least several of these red candidates appear to be bona fide metal-rich GCs (I. Georgiev, private communication).

\subsection{Clustercentric Distribution}
\begin{figure}
\begin{center}
\includegraphics[width=0.5\textwidth]{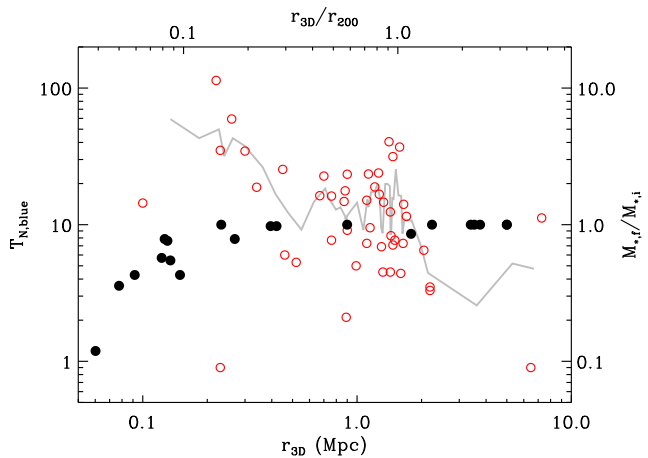}
\caption{Empty circles show \tnblue\ for ACSVCS Virgo dwarfs as a function of their tridimensional distance to M87, while the gray curve represents a smoothed version of the radial trend. Filled circles indicate the final fraction of stellar mass for the harassed galaxies simulated by \citet{Mastropietro2005} as a function of their average clustercentric radius (see text for details). Galaxies that spend longer times deep within the cluster potential are affected the most.  The trend of innermost dEs preferentially having richer GCSs than the outermost ones is the opposite than predicted by mass-stripping mechanisms.
}
\label{fig:gc_radial}
\end{center}
\end{figure}

We have so far focused our discussion on the GCS properties of  Virgo dEs and their potential progenitors in the field. We now turn our attention to what can we learn from  the characteristics of GCSs in early-type dwarfs \emph{within} the cluster. Red circles in Fig.\,\ref{fig:gc_radial} show \tnblue\ for ACSVCS Virgo dEs as a function of their tridimensional distance to M87 (i.e., free from projection effects). As discussed by P08, there is a clear trend for specific frequency to decrease with radius out to $\approx$\,\rvir\ -- except for the two innermost, GC-poor dEs. One could argue that the trend is artificially created by contamination from the GC population of M87, but they show this is unlikely because i) the M87 contribution was accounted for when estimating the final GC specific frequencies, and ii) instead of being randomly distributed, GCs appear to be spatially concentrated around the galaxy body.

In Sect.\,\ref{sect:harass} we argued that harassment is an effective mechanism in a wide range of clustercentric regions, but that it is more efficient where the potential is deeper and in clusters with significant substructure (e.g., \citealt{Gnedin2003}; \citealt{Smith2010}). 
This is also indicated in Fig.\,\ref{fig:gc_radial}, where filled circles represent the fraction of final stellar mass in the harassed dwarfs simulated by \citet{Mastropietro2005} (their table\,1) as a function of their time-averaged cluster radius.\,\footnote{$\langle$r$\rangle=a\,(1-e^{2}/2)$, where all orbits are assumed to be keplerian with semi-major axis $a$ and eccentricity $e$. Note that the simulated dwarfs have been scaled so that the unstripped systems would appear to have \tnblue\ = 10, which roughly corresponds to the median value of actual Virgo dwarfs.} As expected, the longer a galaxy  resides deep within the cluster potential, the higher the mass loss is. 
In this context, the observed negative gradient in \tnblue\ towards the outskirts is puzzling and has in principle no trivial explanation within a framework of late accretion and recent evolution through harassment -- as just the opposite trend is predicted (see also \citealt{Bekki2003}). If mass stripping is more efficient in the inner regions, how is it possible that precisely these dEs contain the richest GCSs? And what causes the GC deficiency towards the outskirts? 
One possibility would be that the outermost dEs have poorer GCSs because they have completed a larger number of orbits (were accreted earlier) within the cluster potential and are just caught at apocentre right now. This is however unlikely, as \citet{Gill2004a} show that older subhaloes experience orbital circularisation and preferentially end up closer to the cluster centre -- and this appears to be consistent with observations too \citep{Lisker2009}.
The two innermost dEs constitute the only exception to the radial trend, and, as discussed in P08, their GC paucity can be naturally explained within a gravitational stripping scenario -- be it through harassment or, most likely, due to the combined tidal fields  of M87, dark matter and hot gas.

A transformation via gas-stripping mechanisms seems even more problematic: why would progenitor dIrrs with more GCs per unit stellar mass have preferentially ended up close to the cluster centre?
An intriguing possibility is that a richer GCS is actually indicative of a higher DM-halo mass \citep{Spitler2009}, and we are just witnessing a \emph{total} mass segregation effect.

\section{Discussion and future prospects}

In the previous Sections we have shown that GCSs can efficiently constrain the role of the different physical mechanisms that operate in clusters and are supposed to shape early-type dwarfs in Virgo. 
The GCSs of low-mass (\mstar\ $< 2\times10^{8}$ \msun) dEs indicate they are consistent with being the remnants of gas-stripped late-type dwarfs, but leave very little room for mechanisms that involve any kind of stellar mass stripping.

The origin of more massive dEs is however rather puzzling. Their very high GC mass specific frequencies represent a challenge for mechanisms that involve the removal of gas from infalling  \emph{field} dwarfs. Moreover, evolution through fast gravitational interactions from luminous progenitors appears to be consistent with only a handful dEs exhibiting the lowest specific frequencies.
Notably, one quarter of the dE(di,bc) subclass exhibit very high GC frequencies (\tnblue\ $>18$), a fact difficult to reconcile with the idea of a late transformation through harassment of accreted disc galaxies. 
This could be unexpected given previous studies suggesting the opposite (e.g., \citealt{Lisker2007}), and these even include evidence from GCSs themselves. 
For instance, \citet{Beasley2009} show that the GCSs of three Virgo dEs (VCC\,1261, VCC\,1087 and VCC\,1528) exhibit significant rotational support, possibly indicating a discy origin for these systems. They propose that a combination of mass loss due to harassment and subsequent stellar population aging can explain the high (v/$\sigma$) of their GCSs, and their compatibility with the Tully-Fisher relation of late-type galaxies. 
Data from the ACSVCS however indicate that VCC\,1261, VCC\,1087 and VCC\,1528 have $T_{N,blue} =$ 6, 19 and 23, respectively. These GC frequencies are 2-7 times larger than the typical values found in field disc galaxies (see Fig.\,\ref{fig:harass_lsb}), so any mass loss scenario would require the progenitors to preferentially lose disc mass (with respect to GCs) by these factors in order to produce comparable remnants -- an unfeasible scenario according to our analysis.

The majority of these massive, GC-rich dEs are however found within the nucleated subclass. These constitute an intriguing population with additional distinct properties, including the oldest stellar population ages, almost negligible rotational support, the roundest shapes and a more concentrated clustercentric distribution (see \citealt{Lisker2009b} and references therein). They comprise a significant fraction of all dEs ($\sim$\,50-75\%; \citealt{Cote2006,Lisker2007}), and are traditionally thought to have formed along with the cluster. 
Having spent a considerable amount of time within the harsh cluster potential, it is  remarkable that these systems have managed to retain such populous GCSs. This is most striking for the population residing in the inner cluster region, where galaxies are supposed to move on circular orbits suffering from substantial dynamical heating \citep{Lisker2009}. The fact that dE(N) contain numerous and spatially extended GCs down to $\approx$\,0.2 Mpc from M87 could be an indication that these systems are shielded by a rather massive DM halo (e.g., \citealt{Penny2009}) that has, so far, prevented the GCS disruption.
In any case, their unique properties, including those of their GCSs, are totally inconsistent with an environmentally-driven \emph{late} origin.

A biased formation scenario for Virgo dEs is however appealing in many aspects. \citet{Diemand2005} show that the hierarchical assembly of DM haloes does not erase the memory of the initial conditions. As a result, the current spatial distribution and kinematics of surviving subhaloes essentially depend on the rarity of the primordial density fluctuations from which they formed. They find that rarer peaks --that depend on both the local and the underlying large-scale fluctuation amplitudes-- collapse earlier and evolve towards a more strongly concentrated configuration within the final halo.
\citet{Moore2006} suggest that protogalaxies and metal-poor globular clusters form out of the first rarer peaks above a critical mass threshold for gas cooling.
The resulting intense star formation activity led to reionisation, delaying the formation of subsequent baryonic structures -- and suppressing it in the lowest mass haloes.
Metal-poor GCs are believed to form in regions with high star formation densities and rates, and in direct proportion to the (early) baryonic content of the (proto)galaxy \citep{McLaughlin1999b} -- which is in turn related to the (sub)halo mass.
As discussed by \citet{Peng2008}, in this scenario "GC formation is biased toward the earliest collapsing halos that can create a large fraction of their stars at high efficiency before reionisation".

Remarkably, this picture provides a (qualitatively) accurate description of the bulk of observed properties of Virgo dEs.
The existence of a morphology-density relation within dEs \citep{Lisker2007} is a natural outcome of this scenario. Indeed, the dE(di,bc)-dE(nN)-dE(N) sequence can be thought of as originating from progressively rarer density peaks, thus explaining the more concentrated (extended) spatial distribution and prevalence of circular (radial) orbits of nucleated (non-nucleated and disc-like) dEs \citep[e.g.,][]{Conselice2001,Lisker2007}.
Moreover, \citet{Peng2008} show that the negative radial gradient in $T_{N}$ for Virgo dEs can be qualitatively explained in terms of a biased GC formation scenario \citep{West1993}. Thus, the inner dEs would be systems that formed earlier, having higher star formation rates (SFR) and surface densities than dwarfs in the outskirts. This would in turn result in a higher efficiency of GC formation, giving rise to the observed radial trend. 
Interestingly, high SFRs are also accompanied by energy release in terms of feedback from supernovae and stellar winds, which can create thick stellar systems \citep{rsj2010} and an extended GC distribution \citep{Mashchenko2008}, as observed in most Virgo dEs. Additionally, gas heating from these feedback mechanisms will make this component more susceptible to subsequent stripping even in the early stages of cluster formation or during group pre-processing (e.g., \citealt{Cortese2006}). 
It is important to note that the majority of these dEs are too massive for feedback processes to completely expel their gas content, and environmental effects are therefore needed to terminate the star formation activity and migrate the dwarf to the red sequence.\,\footnote{The sole existence of the star-forming dE(bc) subclass rules out the former possibility, while the relevant role of environment is reinforced by the existence of a morphology-density relation within dEs \citep{Lisker2007}.}
In this sense, our analysis does not favour violent evolutionary mechanisms that result in substantial stellar mass losses, but more gentle processes involving gas removal by a combination of internal and external factors \citep[e.g.,][]{Boselli2008}.
The tidal stirring scenario \citep{Mayer2001a,Kazantzidis2011} provides a plausible transformation mechanism, as long as GC and stellar mass losses are not prominent.

The actual importance of stellar mass loss in cluster galaxy evolution can be quantified through the study of intra-cluster globular clusters (IGCs). It is now well established that a significant population of such unbound systems inhabit the intergalactic regions of relatively massive clusters \citep{Lee2010,Peng2011,West2011}. It is however not yet clear what fraction of these IGCs originate from tidal interactions and what fraction constitute a primordial cluster population \citep{West1995}. In particular, spectroscopic determinations of velocities and metallicities for these systems will prove critical to address their dynamical status and their connection with cluster galaxies GCSs \citep[e.g.,][]{Romanowsky2012}.

We finally stress that the analysis of GCSs suggests that the progenitors of Virgo dEs are intrinsically different from today's \emph{field} dwarf galaxies -- but \emph{cluster} dIrrs appear to be fully consistent.
Interestingly, the significantly different metal-poor GC frequencies of these two late-type dwarfs can also be naturally explained within the biased formation scenario.
Virgo dIrr galaxies collapsed earlier from higher-$\sigma$ peaks than their counterparts in low-density  environments, thus forming a larger fraction of their current stellar mass in the form of GCs before reionisation. Star formation was resumed shortly after this event, but under very different conditions for both systems: while cluster dIrrs had to struggle within the developing harsh potential well, field dwarfs have been able to form stars without significant distress until today.
The difference in \tnblue\ further suggests that, modulo internal evolutionary effects in the GCS,  the initial star formation histories of cluster and field dwarfs must also differ.
This nothing but reinforces the  conclusion of \citet{Skillman1995} that "present day dEs and dIs may share a common ancestor, but dEs do not evolve from dIs".

In this study we have shown that GCSs are very sensitive tracers of evolutionary mechanisms in high-density environments, and thus put strong constraints on the relevance of the various processes. In forthcoming papers we will present a more elaborated treatment of the effects of fast interactions on the GCSs of early- and late-type galaxies using numerical simulations, which will further allow us to compare their kinematics with available observations. 
Despite of the growing amount of data, our understanding of GCSs in Virgo dEs is still rather limited. 
The situation will dramatically change once the Next Generation Virgo Survey (Ferrarese et al. 2012) is completed. The analysis of GCSs in a complete sample of Virgo dwarfs as a function of their stellar mass, morphological type and clustercentric position should provide us with a clear picture of their origin and the role of environmental effects in shaping their properties.

\section*{Acknowledgements}
The authors acknowledge an anonymous referee for helpful feedback and suggestions that improved the presentation of results. We benefited from fruitful discussions with T. Puzia, J. Pe\~narrubia, C. Gonz\'alez-Garc\'ia, I. Georgiev and L. Cortese. RSJ thanks M. West, R. Smith and D. Forbes for many useful comments on an earlier version of the manuscript, and J. Cenarro for driving his attention towards this topic. 
This work was co-funded under the Marie Curie Actions of the European Commission (FP7-COFUND).
JALA was supported by the projects AYA2010-21887-C04-04 and by the Consolider-Ingenio 2010 Program grant CSD2006-00070.

\bibliographystyle{mn2e}
\bibliography{/Users/rsanchez/WORK/PAPERS/rsj_references.bib}

\begin{thebibliography}{}

\bibitem[\protect\citeauthoryear{{Aguerri} \&
  {Gonz{\'a}lez-Garc{\'{\i}}a}}{{Aguerri} \&
  {Gonz{\'a}lez-Garc{\'{\i}}a}}{2009}]{Aguerri2009a}
{Aguerri} J.~A.~L.,  {Gonz{\'a}lez-Garc{\'{\i}}a} A.~C.,  2009, \aap, 494, 891

\bibitem[\protect\citeauthoryear{{Aguerri}, {Iglesias-P{\'a}ramo},
  {V{\'{\i}}lchez}, {Mu{\~n}oz-Tu{\~n}{\'o}n} \&
  {S{\'a}nchez-Janssen}}{{Aguerri} et~al.}{2005}]{Aguerri2005}
{Aguerri} J.~A.~L.,  {Iglesias-P{\'a}ramo} J.,  {V{\'{\i}}lchez} J.~M.,
  {Mu{\~n}oz-Tu{\~n}{\'o}n} C.,    {S{\'a}nchez-Janssen} R.,  2005, \aj, 130,
  475

\bibitem[\protect\citeauthoryear{{Aguilar} \& {White}}{{Aguilar} \&
  {White}}{1985}]{Aguilar1985}
{Aguilar} L.~A.,  {White} S.~D.~M.,  1985, \apj, 295, 374

\bibitem[\protect\citeauthoryear{{Arag{\'o}n-Salamanca}, {Bedregal} \&
  {Merrifield}}{{Arag{\'o}n-Salamanca} et~al.}{2006}]{AragonSalamanca2006}
{Arag{\'o}n-Salamanca} A.,  {Bedregal} A.~G.,    {Merrifield} M.~R.,  2006,
  \aap, 458, 101

\bibitem[\protect\citeauthoryear{{Barazza}, {Binggeli} \& {Jerjen}}{{Barazza}
  et~al.}{2002}]{Barazza2002}
{Barazza} F.~D.,  {Binggeli} B.,    {Jerjen} H.,  2002, \aap, 391, 823

\bibitem[\protect\citeauthoryear{{Beasley}, {Baugh}, {Forbes}, {Sharples} \&
  {Frenk}}{{Beasley} et~al.}{2002}]{Beasley2002}
{Beasley} M.~A.,  {Baugh} C.~M.,  {Forbes} D.~A.,  {Sharples} R.~M.,    {Frenk}
  C.~S.,  2002, \mnras, 333, 383

\bibitem[\protect\citeauthoryear{{Beasley}, {Cenarro}, {Strader} \&
  {Brodie}}{{Beasley} et~al.}{2009}]{Beasley2009}
{Beasley} M.~A.,  {Cenarro} A.~J.,  {Strader} J.,    {Brodie} J.~P.,  2009,
  \aj, 137, 5146

\bibitem[\protect\citeauthoryear{{Beasley}, {Strader}, {Brodie}, {Cenarro} \&
  {Geha}}{{Beasley} et~al.}{2006}]{Beasley2006}
{Beasley} M.~A.,  {Strader} J.,  {Brodie} J.~P.,  {Cenarro} A.~J.,    {Geha}
  M.,  2006, \aj, 131, 814

\bibitem[\protect\citeauthoryear{{Bekki}, {Forbes}, {Beasley} \&
  {Couch}}{{Bekki} et~al.}{2003}]{Bekki2003}
{Bekki} K.,  {Forbes} D.~A.,  {Beasley} M.~A.,    {Couch} W.~J.,  2003, \mnras,
  344, 1334

\bibitem[\protect\citeauthoryear{{Bica}, {Bonatto}, {Barbuy} \&
  {Ortolani}}{{Bica} et~al.}{2006}]{Bica2006}
{Bica} E.,  {Bonatto} C.,  {Barbuy} B.,    {Ortolani} S.,  2006, \aap, 450, 105

\bibitem[\protect\citeauthoryear{{Binggeli}, {Sandage} \& {Tammann}}{{Binggeli}
  et~al.}{1985}]{Binggeli1985}
{Binggeli} B.,  {Sandage} A.,    {Tammann} G.~A.,  1985, \aj, 90, 1681

\bibitem[\protect\citeauthoryear{{Boselli}, {Boissier}, {Cortese} \&
  {Gavazzi}}{{Boselli} et~al.}{2008}]{Boselli2008}
{Boselli} A.,  {Boissier} S.,  {Cortese} L.,    {Gavazzi} G.,  2008, \apj, 674,
  742

\bibitem[\protect\citeauthoryear{{Boselli} \& {Gavazzi}}{{Boselli} \&
  {Gavazzi}}{2006}]{Boselli2006}
{Boselli} A.,  {Gavazzi} G.,  2006, \pasp, 118, 517

\bibitem[\protect\citeauthoryear{{Brodie} \& {Strader}}{{Brodie} \&
  {Strader}}{2006}]{Brodie2006}
{Brodie} J.~P.,  {Strader} J.,  2006, \araa, 44, 193

\bibitem[\protect\citeauthoryear{{Chandar}, {Whitmore} \& {Lee}}{{Chandar}
  et~al.}{2004}]{Chandar2004}
{Chandar} R.,  {Whitmore} B.,    {Lee} M.~G.,  2004, \apj, 611, 220

\bibitem[\protect\citeauthoryear{{Coenda}, {Muriel} \& {Donzelli}}{{Coenda}
  et~al.}{2009}]{Coenda2009}
{Coenda} V.,  {Muriel} H.,    {Donzelli} C.,  2009, \apj, 700, 1382

\bibitem[\protect\citeauthoryear{{Cohen}, {Blakeslee} \& {Ryzhov}}{{Cohen}
  et~al.}{1998}]{Cohen1998}
{Cohen} J.~G.,  {Blakeslee} J.~P.,    {Ryzhov} A.,  1998, \apj, 496, 808

\bibitem[\protect\citeauthoryear{{Conselice}, {Gallagher} III \&
  {Wyse}}{{Conselice} et~al.}{2001}]{Conselice2001}
{Conselice} C.~J.,  {Gallagher} III J.~S.,    {Wyse} R.~F.~G.,  2001, \apj,
  559, 791

\bibitem[\protect\citeauthoryear{{Cortese}, {Gavazzi}, {Boselli}, {Franzetti},
  {Kennicutt}, {O'Neil} \& {Sakai}}{{Cortese} et~al.}{2006}]{Cortese2006}
{Cortese} L.,  {Gavazzi} G.,  {Boselli} A.,  {Franzetti} P.,  {Kennicutt}
  R.~C.,  {O'Neil} K.,    {Sakai} S.,  2006, \aap, 453, 847

\bibitem[\protect\citeauthoryear{{C{\^o}t{\'e}}, {Blakeslee}, {Ferrarese},
  {Jord{\'a}n}, {Mei}, {Merritt}, {Milosavljevi{\'c}}, {Peng}, {Tonry} \&
  {West}}{{C{\^o}t{\'e}} et~al.}{2004}]{Cote2004}
{C{\^o}t{\'e}} P.,  {Blakeslee} J.~P.,  {Ferrarese} L.,  {Jord{\'a}n} A.,
  {Mei} S.,  {Merritt} D.,  {Milosavljevi{\'c}} M.,  {Peng} E.~W.,  {Tonry}
  J.~L.,    {West} M.~J.,  2004, \apjs, 153, 223

\bibitem[\protect\citeauthoryear{{C{\^o}t{\'e}}, {Piatek}, {Ferrarese},
  {Jord{\'a}n}, {Merritt}, {Peng}, {Ha{\c s}egan}, {Blakeslee}, {Mei}, {West},
  {Milosavljevi{\'c}} \& {Tonry}}{{C{\^o}t{\'e}} et~al.}{2006}]{Cote2006}
{C{\^o}t{\'e}} P.,  {Piatek} S.,  {Ferrarese} L.,  {Jord{\'a}n} A.,  {Merritt}
  D.,  {Peng} E.~W.,  {Ha{\c s}egan} M.,  {Blakeslee} J.~P.,  {Mei} S.,  {West}
  M.~J.,  {Milosavljevi{\'c}} M.,    {Tonry} J.~L.,  2006, \apjs, 165, 57

\bibitem[\protect\citeauthoryear{{Cowie} \& {Songaila}}{{Cowie} \&
  {Songaila}}{1977}]{Cowie1977}
{Cowie} L.~L.,  {Songaila} A.,  1977, \nat, 266, 501

\bibitem[\protect\citeauthoryear{{De Lucia}, {Poggianti},
  {Arag{\'o}n-Salamanca}, {White}, {Zaritsky}, {Clowe}, {Halliday}, {Jablonka},
  {von der Linden}, {Milvang-Jensen}, {Pell{\'o}}, {Rudnick}, {Saglia} \&
  {Simard}}{{De Lucia} et~al.}{2007}]{DeLucia2007}
{De Lucia} G.,  {Poggianti} B.~M.,  {Arag{\'o}n-Salamanca} A.,  {White}
  S.~D.~M.,  {Zaritsky} D.,  {Clowe} D.,  {Halliday} C.,  {Jablonka} P.,  {von
  der Linden} A.,  {Milvang-Jensen} B.,  {Pell{\'o}} R.,  {Rudnick} G.,
  {Saglia} R.~P.,    {Simard} L.,  2007, \mnras, 374, 809

\bibitem[\protect\citeauthoryear{{de Rijcke}, {Michielsen}, {Dejonghe},
  {Zeilinger} \& {Hau}}{{de Rijcke} et~al.}{2005}]{deRijcke2005}
{de Rijcke} S.,  {Michielsen} D.,  {Dejonghe} H.,  {Zeilinger} W.~W.,    {Hau}
  G.~K.~T.,  2005, \aap, 438, 491

\bibitem[\protect\citeauthoryear{{Dekel} \& {Silk}}{{Dekel} \&
  {Silk}}{1986}]{Dekel1986}
{Dekel} A.,  {Silk} J.,  1986, \apj, 303, 39

\bibitem[\protect\citeauthoryear{{Diemand}, {Madau} \& {Moore}}{{Diemand}
  et~al.}{2005}]{Diemand2005}
{Diemand} J.,  {Madau} P.,    {Moore} B.,  2005, \mnras, 364, 367

\bibitem[\protect\citeauthoryear{{Drinkwater}, {Gregg} \&
  {Colless}}{{Drinkwater} et~al.}{2001}]{Drinkwater2001}
{Drinkwater} M.~J.,  {Gregg} M.~D.,    {Colless} M.,  2001, \apjl, 548, L139

\bibitem[\protect\citeauthoryear{{Ferrarese}, {C{\^o}t{\'e}}, {Jord{\'a}n},
  {Peng}, {Blakeslee}, {Piatek}, {Mei}, {Merritt}, {Milosavljevi{\'c}}, {Tonry}
  \& {West}}{{Ferrarese} et~al.}{2006}]{Ferrarese2006}
{Ferrarese} L.,  {C{\^o}t{\'e}} P.,  {Jord{\'a}n} A.,  {Peng} E.~W.,
  {Blakeslee} J.~P.,  {Piatek} S.,  {Mei} S.,  {Merritt} D.,
  {Milosavljevi{\'c}} M.,  {Tonry} J.~L.,    {West} M.~J.,  2006, \apjs, 164,
  334

\bibitem[\protect\citeauthoryear{{Forbes}, {Brodie} \& {Larsen}}{{Forbes}
  et~al.}{2001}]{Forbes2001}
{Forbes} D.~A.,  {Brodie} J.~P.,    {Larsen} S.~S.,  2001, \apjl, 556, L83

\bibitem[\protect\citeauthoryear{{Gadotti} \& {S{\'a}nchez-Janssen}}{{Gadotti}
  \& {S{\'a}nchez-Janssen}}{2011}]{GadottiSJ2012}
{Gadotti} D.~A.,  {S{\'a}nchez-Janssen} R.,  2011, ArXiv e-prints

\bibitem[\protect\citeauthoryear{{Geha}, {Guhathakurta} \& {van der
  Marel}}{{Geha} et~al.}{2003}]{Geha2003}
{Geha} M.,  {Guhathakurta} P.,    {van der Marel} R.~P.,  2003, \aj, 126, 1794

\bibitem[\protect\citeauthoryear{{Georgiev}, {Puzia}, {Goudfrooij} \&
  {Hilker}}{{Georgiev} et~al.}{2010}]{Georgiev2010}
{Georgiev} I.~Y.,  {Puzia} T.~H.,  {Goudfrooij} P.,    {Hilker} M.,  2010,
  \mnras, 406, 1967

\bibitem[\protect\citeauthoryear{{Gill}, {Knebe}, {Gibson} \& {Dopita}}{{Gill}
  et~al.}{2004}]{Gill2004a}
{Gill} S.~P.~D.,  {Knebe} A.,  {Gibson} B.~K.,    {Dopita} M.~A.,  2004,
  \mnras, 351, 410

\bibitem[\protect\citeauthoryear{{Gnedin}}{{Gnedin}}{2003}]{Gnedin2003}
{Gnedin} O.~Y.,  2003, \apj, 582, 141

\bibitem[\protect\citeauthoryear{{Gnedin}, {Hernquist} \& {Ostriker}}{{Gnedin}
  et~al.}{1999}]{Gnedin1999a}
{Gnedin} O.~Y.,  {Hernquist} L.,    {Ostriker} J.~P.,  1999, \apj, 514, 109

\bibitem[\protect\citeauthoryear{{Goudfrooij}, {Strader}, {Brenneman},
  {Kissler-Patig}, {Minniti} \& {Edwin Huizinga}}{{Goudfrooij}
  et~al.}{2003}]{Goudfrooij2003}
{Goudfrooij} P.,  {Strader} J.,  {Brenneman} L.,  {Kissler-Patig} M.,
  {Minniti} D.,    {Edwin Huizinga} J.,  2003, \mnras, 343, 665

\bibitem[\protect\citeauthoryear{{Governato}, {Brook}, {Mayer} \& {et
  al.}}{{Governato} et~al.}{2010}]{Governato2010}
{Governato} F.,  {Brook} C.,  {Mayer} L.,    {et al.} 2010, \nat, 463, 203

\bibitem[\protect\citeauthoryear{{Graham} \& {Guzm{\'a}n}}{{Graham} \&
  {Guzm{\'a}n}}{2003}]{Graham2003}
{Graham} A.~W.,  {Guzm{\'a}n} R.,  2003, \aj, 125, 2936

\bibitem[\protect\citeauthoryear{{Gunn} \& {Gott}}{{Gunn} \&
  {Gott}}{1972}]{GunnGott1972}
{Gunn} J.~E.,  {Gott} J.~R.~I.,  1972, \apj, 176, 1

\bibitem[\protect\citeauthoryear{{Haines}, {La Barbera}, {Mercurio}, {Merluzzi}
  \& {Busarello}}{{Haines} et~al.}{2006}]{Haines2006}
{Haines} C.~P.,  {La Barbera} F.,  {Mercurio} A.,  {Merluzzi} P.,
  {Busarello} G.,  2006, \apjl, 647, L21

\bibitem[\protect\citeauthoryear{{Harris} \& {van den Bergh}}{{Harris} \& {van
  den Bergh}}{1981}]{Harris1981}
{Harris} W.~E.,  {van den Bergh} S.,  1981, \aj, 86, 1627

\bibitem[\protect\citeauthoryear{{Janz}, {Laurikainen}, {Lisker}, {Salo},
  {Peletier}, {Niemi}, {den Brok}, {Toloba}, {Falc{\'o}n-Barroso}, {Boselli} \&
  {Hensler}}{{Janz} et~al.}{2012}]{Janz2012}
{Janz} J.,  {Laurikainen} E.,  {Lisker} T.,  {Salo} H.,  {Peletier} R.~F.,
  {Niemi} S.-M.,  {den Brok} M.,  {Toloba} E.,  {Falc{\'o}n-Barroso} J.,
  {Boselli} A.,    {Hensler} G.,  2012, \apjl, 745, L24

\bibitem[\protect\citeauthoryear{{Jerjen}, {Kalnajs} \& {Binggeli}}{{Jerjen}
  et~al.}{2000}]{Jerjen2000}
{Jerjen} H.,  {Kalnajs} A.,    {Binggeli} B.,  2000, \aap, 358, 845

\bibitem[\protect\citeauthoryear{{Kazantzidis}, {{\L}okas}, {Callegari},
  {Mayer} \& {Moustakas}}{{Kazantzidis} et~al.}{2011}]{Kazantzidis2011}
{Kazantzidis} S.,  {{\L}okas} E.~L.,  {Callegari} S.,  {Mayer} L.,
  {Moustakas} L.~A.,  2011, \apj, 726, 98

\bibitem[\protect\citeauthoryear{{Kormendy} \& {Bender}}{{Kormendy} \&
  {Bender}}{2012}]{Kormendy2012}
{Kormendy} J.,  {Bender} R.,  2012, \apjs, 198, 2

\bibitem[\protect\citeauthoryear{{Kormendy}, {Fisher}, {Cornell} \&
  {Bender}}{{Kormendy} et~al.}{2009}]{Kormendy2009}
{Kormendy} J.,  {Fisher} D.~B.,  {Cornell} M.~E.,    {Bender} R.,  2009, \apjs,
  182, 216

\bibitem[\protect\citeauthoryear{{Larson}, {Tinsley} \& {Caldwell}}{{Larson}
  et~al.}{1980}]{Larson1980}
{Larson} R.~B.,  {Tinsley} B.~M.,    {Caldwell} C.~N.,  1980, \apj, 237, 692

\bibitem[\protect\citeauthoryear{{Lee}, {Park} \& {Hwang}}{{Lee}
  et~al.}{2010}]{Lee2010}
{Lee} M.~G.,  {Park} H.~S.,    {Hwang} H.~S.,  2010, Science, 328, 334

\bibitem[\protect\citeauthoryear{{Lisker}}{{Lisker}}{2009}]{Lisker2009b}
{Lisker} T.,  2009, Astronomische Nachrichten, 330, 1043

\bibitem[\protect\citeauthoryear{{Lisker}, {Grebel} \& {Binggeli}}{{Lisker}
  et~al.}{2006}]{Lisker2006}
{Lisker} T.,  {Grebel} E.~K.,    {Binggeli} B.,  2006, \aj, 132, 497

\bibitem[\protect\citeauthoryear{{Lisker}, {Grebel}, {Binggeli} \&
  {Glatt}}{{Lisker} et~al.}{2007}]{Lisker2007}
{Lisker} T.,  {Grebel} E.~K.,  {Binggeli} B.,    {Glatt} K.,  2007, \apj, 660,
  1186

\bibitem[\protect\citeauthoryear{{Lisker}, {Janz}, {Hensler}, {Kim}, {Rey},
  {Weinmann}, {Mastropietro}, {Hielscher}, {Paudel} \& {Kotulla}}{{Lisker}
  et~al.}{2009}]{Lisker2009}
{Lisker} T.,  {Janz} J.,  {Hensler} G.,  {Kim} S.,  {Rey} S.,  {Weinmann} S.,
  {Mastropietro} C.,  {Hielscher} O.,  {Paudel} S.,    {Kotulla} R.,  2009,
  \apjl, 706, L124

\bibitem[\protect\citeauthoryear{{Mashchenko}, {Wadsley} \&
  {Couchman}}{{Mashchenko} et~al.}{2008}]{Mashchenko2008}
{Mashchenko} S.,  {Wadsley} J.,    {Couchman} H.~M.~P.,  2008, Science, 319,
  174

\bibitem[\protect\citeauthoryear{{Mastropietro}, {Moore}, {Mayer},
  {Debattista}, {Piffaretti} \& {Stadel}}{{Mastropietro}
  et~al.}{2005}]{Mastropietro2005}
{Mastropietro} C.,  {Moore} B.,  {Mayer} L.,  {Debattista} V.~P.,  {Piffaretti}
  R.,    {Stadel} J.,  2005, \mnras, 364, 607

\bibitem[\protect\citeauthoryear{{Mayer}, {Governato}, {Colpi}, {Moore},
  {Quinn}, {Wadsley}, {Stadel} \& {Lake}}{{Mayer} et~al.}{2001}]{Mayer2001a}
{Mayer} L.,  {Governato} F.,  {Colpi} M.,  {Moore} B.,  {Quinn} T.,  {Wadsley}
  J.,  {Stadel} J.,    {Lake} G.,  2001, \apj, 559, 754

\bibitem[\protect\citeauthoryear{{McLaughlin}}{{McLaughlin}}{1999}]{McLaughlin%
1999b}
{McLaughlin} D.~E.,  1999, \aj, 117, 2398

\bibitem[\protect\citeauthoryear{{Mei}, {Blakeslee}, {C{\^o}t{\'e}}, {Tonry},
  {West}, {Ferrarese}, {Jord{\'a}n}, {Peng}, {Anthony} \& {Merritt}}{{Mei}
  et~al.}{2007}]{Mei2007}
{Mei} S.,  {Blakeslee} J.~P.,  {C{\^o}t{\'e}} P.,  {Tonry} J.~L.,  {West}
  M.~J.,  {Ferrarese} L.,  {Jord{\'a}n} A.,  {Peng} E.~W.,  {Anthony} A.,
  {Merritt} D.,  2007, \apj, 655, 144

\bibitem[\protect\citeauthoryear{{M{\'e}ndez-Abreu}, {S{\'a}nchez-Janssen} \&
  {Aguerri}}{{M{\'e}ndez-Abreu} et~al.}{2010}]{Mendez-Abreu2010}
{M{\'e}ndez-Abreu} J.,  {S{\'a}nchez-Janssen} R.,    {Aguerri} J.~A.~L.,  2010,
  \apjl, 711, L61

\bibitem[\protect\citeauthoryear{{Merritt}}{{Merritt}}{1984}]{Merritt1984}
{Merritt} D.,  1984, \apj, 276, 26

\bibitem[\protect\citeauthoryear{{Miller} \& {Lotz}}{{Miller} \&
  {Lotz}}{2007}]{Miller2007}
{Miller} B.~W.,  {Lotz} J.~M.,  2007, \apj, 670, 1074

\bibitem[\protect\citeauthoryear{{Moore}, {Diemand}, {Madau}, {Zemp} \&
  {Stadel}}{{Moore} et~al.}{2006}]{Moore2006}
{Moore} B.,  {Diemand} J.,  {Madau} P.,  {Zemp} M.,    {Stadel} J.,  2006,
  \mnras, 368, 563

\bibitem[\protect\citeauthoryear{{Moore}, {Katz}, {Lake}, {Dressler} \&
  {Oemler}}{{Moore} et~al.}{1996}]{Moore1996}
{Moore} B.,  {Katz} N.,  {Lake} G.,  {Dressler} A.,    {Oemler} A.,  1996,
  \nat, 379, 613

\bibitem[\protect\citeauthoryear{{Moore}, {Lake} \& {Katz}}{{Moore}
  et~al.}{1998}]{Moore1998}
{Moore} B.,  {Lake} G.,    {Katz} N.,  1998, \apj, 495, 139

\bibitem[\protect\citeauthoryear{{Moore}, {Lake}, {Quinn} \& {Stadel}}{{Moore}
  et~al.}{1999}]{Moore1999}
{Moore} B.,  {Lake} G.,  {Quinn} T.,    {Stadel} J.,  1999, \mnras, 304, 465

\bibitem[\protect\citeauthoryear{{Mora}, {Larsen} \& {Kissler-Patig}}{{Mora}
  et~al.}{2007}]{Mora2007}
{Mora} M.~D.,  {Larsen} S.~S.,    {Kissler-Patig} M.,  2007, \aap, 464, 495

\bibitem[\protect\citeauthoryear{{Nulsen}}{{Nulsen}}{1982}]{Nulsen1982}
{Nulsen} P.~E.~J.,  1982, \mnras, 198, 1007

\bibitem[\protect\citeauthoryear{{Oh} \& {Lin}}{{Oh} \& {Lin}}{2000}]{Oh2000}
{Oh} K.~S.,  {Lin} D.~N.~C.,  2000, \apj, 543, 620

\bibitem[\protect\citeauthoryear{{Pedraz}, {Gorgas}, {Cardiel},
  {S{\'a}nchez-Bl{\'a}zquez} \& {Guzm{\'a}n}}{{Pedraz}
  et~al.}{2002}]{Pedraz2002}
{Pedraz} S.,  {Gorgas} J.,  {Cardiel} N.,  {S{\'a}nchez-Bl{\'a}zquez} P.,
  {Guzm{\'a}n} R.,  2002, \mnras, 332, L59

\bibitem[\protect\citeauthoryear{{Peng}, {Ferguson}, {Goudfrooij} \& {et
  al.}}{{Peng} et~al.}{2011}]{Peng2011}
{Peng} E.~W.,  {Ferguson} H.~C.,  {Goudfrooij} P.,    {et al.} 2011, \apj, 730,
  23

\bibitem[\protect\citeauthoryear{{Peng}, {Jord{\'a}n}, {C{\^o}t{\'e}},
  {Blakeslee}, {Ferrarese}, {Mei}, {West}, {Merritt}, {Milosavljevi{\'c}} \&
  {Tonry}}{{Peng} et~al.}{2006}]{Peng2006}
{Peng} E.~W.,  {Jord{\'a}n} A.,  {C{\^o}t{\'e}} P.,  {Blakeslee} J.~P.,
  {Ferrarese} L.,  {Mei} S.,  {West} M.~J.,  {Merritt} D.,  {Milosavljevi{\'c}}
  M.,    {Tonry} J.~L.,  2006, \apj, 639, 95

\bibitem[\protect\citeauthoryear{{Peng}, {Jord{\'a}n}, {C{\^o}t{\'e}},
  {Takamiya}, {West}, {Blakeslee}, {Chen}, {Ferrarese}, {Mei}, {Tonry} \&
  {West}}{{Peng} et~al.}{2008}]{Peng2008}
{Peng} E.~W.,  {Jord{\'a}n} A.,  {C{\^o}t{\'e}} P.,  {Takamiya} M.,  {West}
  M.~J.,  {Blakeslee} J.~P.,  {Chen} C.-W.,  {Ferrarese} L.,  {Mei} S.,
  {Tonry} J.~L.,    {West} A.~A.,  2008, \apj, 681, 197

\bibitem[\protect\citeauthoryear{{Penny}, {Conselice}, {de Rijcke} \&
  {Held}}{{Penny} et~al.}{2009}]{Penny2009}
{Penny} S.~J.,  {Conselice} C.~J.,  {de Rijcke} S.,    {Held} E.~V.,  2009,
  \mnras, 393, 1054

\bibitem[\protect\citeauthoryear{{Puzia}, {Kissler-Patig}, {Brodie} \&
  {Huchra}}{{Puzia} et~al.}{1999}]{Puzia1999}
{Puzia} T.~H.,  {Kissler-Patig} M.,  {Brodie} J.~P.,    {Huchra} J.~P.,  1999,
  \aj, 118, 2734

\bibitem[\protect\citeauthoryear{{Quilis}, {Moore} \& {Bower}}{{Quilis}
  et~al.}{2000}]{Quilis2000}
{Quilis} V.,  {Moore} B.,    {Bower} R.,  2000, Science, 288, 1617

\bibitem[\protect\citeauthoryear{{Rhode} \& {Zepf}}{{Rhode} \&
  {Zepf}}{2004}]{Rhode2004}
{Rhode} K.~L.,  {Zepf} S.~E.,  2004, \aj, 127, 302

\bibitem[\protect\citeauthoryear{{Rhode}, {Zepf}, {Kundu} \& {Larner}}{{Rhode}
  et~al.}{2007}]{Rhode2007}
{Rhode} K.~L.,  {Zepf} S.~E.,  {Kundu} A.,    {Larner} A.~N.,  2007, \aj, 134,
  1403

\bibitem[\protect\citeauthoryear{{Romanowsky}, {Strader}, {Brodie}, {Mihos},
  {Spitler}, {Forbes}, {Foster} \& {Arnold}}{{Romanowsky}
  et~al.}{2012}]{Romanowsky2012}
{Romanowsky} A.~J.,  {Strader} J.,  {Brodie} J.~P.,  {Mihos} J.~C.,  {Spitler}
  L.~R.,  {Forbes} D.~A.,  {Foster} C.,    {Arnold} J.~A.,  2012, \apj, 748, 29

\bibitem[\protect\citeauthoryear{{S{\'a}nchez-Janssen}, {Aguerri} \&
  {Mu{\~n}oz-Tu{\~n}{\'o}n}}{{S{\'a}nchez-Janssen} et~al.}{2008}]{rsj2008}
{S{\'a}nchez-Janssen} R.,  {Aguerri} J.~A.~L.,    {Mu{\~n}oz-Tu{\~n}{\'o}n} C.,
   2008, \apjl, 679, L77

\bibitem[\protect\citeauthoryear{{S{\'a}nchez-Janssen}, {M{\'e}ndez-Abreu} \&
  {Aguerri}}{{S{\'a}nchez-Janssen} et~al.}{2010}]{rsj2010}
{S{\'a}nchez-Janssen} R.,  {M{\'e}ndez-Abreu} J.,    {Aguerri} J.~A.~L.,  2010,
  \mnras, 406, L65

\bibitem[\protect\citeauthoryear{{Seth}, {Olsen}, {Miller}, {Lotz} \&
  {Telford}}{{Seth} et~al.}{2004}]{Seth2004}
{Seth} A.,  {Olsen} K.,  {Miller} B.,  {Lotz} J.,    {Telford} R.,  2004, \aj,
  127, 798

\bibitem[\protect\citeauthoryear{{Skillman} \& {Bender}}{{Skillman} \&
  {Bender}}{1995}]{Skillman1995}
{Skillman} E.~D.,  {Bender} R.,  1995, in {M.~Pena \& S.~Kurtz} ed., Revista
  Mexicana de Astronomia y Astrofisica Conference Series Vol.~3, {The Dwarf
  Galaxy Star Formation Crisis (Invited paper)}.
p.~25

\bibitem[\protect\citeauthoryear{{Smith}, {Davies} \& {Nelson}}{{Smith}
  et~al.}{2010}]{Smith2010}
{Smith} R.,  {Davies} J.~I.,    {Nelson} A.~H.,  2010, \mnras, 405, 1723

\bibitem[\protect\citeauthoryear{{Smith}, {Lucey}, {Hudson}, {Allanson},
  {Bridges}, {Hornschemeier}, {Marzke} \& {Miller}}{{Smith}
  et~al.}{2009}]{Smith2009}
{Smith} R.~J.,  {Lucey} J.~R.,  {Hudson} M.~J.,  {Allanson} S.~P.,  {Bridges}
  T.~J.,  {Hornschemeier} A.~E.,  {Marzke} R.~O.,    {Miller} N.~A.,  2009,
  \mnras, 392, 1265

\bibitem[\protect\citeauthoryear{{Spitler} \& {Forbes}}{{Spitler} \&
  {Forbes}}{2009}]{Spitler2009}
{Spitler} L.~R.,  {Forbes} D.~A.,  2009, \mnras, 392, L1

\bibitem[\protect\citeauthoryear{{Spitler}, {Forbes}, {Strader}, {Brodie} \&
  {Gallagher}}{{Spitler} et~al.}{2008}]{Spitler2008}
{Spitler} L.~R.,  {Forbes} D.~A.,  {Strader} J.,  {Brodie} J.~P.,
  {Gallagher} J.~S.,  2008, \mnras, 385, 361

\bibitem[\protect\citeauthoryear{{Spitzer} Jr.}{{Spitzer}}{1958}]{Spitzer1958}
{Spitzer} Jr. L.,  1958, \apj, 127, 17

\bibitem[\protect\citeauthoryear{{Toloba}, {Boselli}, {Cenarro}, {Peletier},
  {Gorgas}, {Gil de Paz} \& {Mu{\~n}oz-Mateos}}{{Toloba}
  et~al.}{2011}]{Toloba2011}
{Toloba} E.,  {Boselli} A.,  {Cenarro} A.~J.,  {Peletier} R.~F.,  {Gorgas} J.,
  {Gil de Paz} A.,    {Mu{\~n}oz-Mateos} J.~C.,  2011, \aap, 526, A114

\bibitem[\protect\citeauthoryear{{van Zee}, {Skillman} \& {Haynes}}{{van Zee}
  et~al.}{2004}]{vanZee2004}
{van Zee} L.,  {Skillman} E.~D.,    {Haynes} M.~P.,  2004, \aj, 128, 121

\bibitem[\protect\citeauthoryear{{Villegas}, {Kissler-Patig}, {Jord{\'a}n},
  {Goudfrooij} \& {Zwaan}}{{Villegas} et~al.}{2008}]{Villegas2008}
{Villegas} D.,  {Kissler-Patig} M.,  {Jord{\'a}n} A.,  {Goudfrooij} P.,
  {Zwaan} M.,  2008, \aj, 135, 467

\bibitem[\protect\citeauthoryear{{West}}{{West}}{1993}]{West1993}
{West} M.~J.,  1993, \mnras, 265, 755

\bibitem[\protect\citeauthoryear{{West}, {Cote}, {Jones}, {Forman} \&
  {Marzke}}{{West} et~al.}{1995}]{West1995}
{West} M.~J.,  {Cote} P.,  {Jones} C.,  {Forman} W.,    {Marzke} R.~O.,  1995,
  \apjl, 453, L77+

\bibitem[\protect\citeauthoryear{{West}, {Jordan}, {Blakeslee}, {Cote},
  {Gregg}, {Takamiya} \& {Marzke}}{{West} et~al.}{2011}]{West2011}
{West} M.~J.,  {Jordan} A.,  {Blakeslee} J.~P.,  {Cote} P.,  {Gregg} M.~D.,
  {Takamiya} M.,    {Marzke} R.~O.,  2011, ArXiv e-prints

\bibitem[\protect\citeauthoryear{{Wilman} \& {Erwin}}{{Wilman} \&
  {Erwin}}{2012}]{Wilman2012}
{Wilman} D.~J.,  {Erwin} P.,  2012, \apj, 746, 160

\bibitem[\protect\citeauthoryear{{Zepf} \& {Ashman}}{{Zepf} \&
  {Ashman}}{1993}]{Zepf1993}
{Zepf} S.~E.,  {Ashman} K.~M.,  1993, \mnras, 264, 611

\end{thebibliography}

\appendix
\section{A note on the origin of cluster S0 galaxies}
\label{appendix}
Figures\,\ref{fig:harass_lsb} and \ref{fig:harass_fred_evol} show that S0 and Sp galaxies share similar locations in both the \mstar-\tnblue\ and \mstar-red GC fraction planes. Taken at face value, this would argue in favour of an evolutionary link between both galaxy types, as environmental effects probably do not change significantly the quantities involved: ram-pressure stripping does not affect the stellar components \citep{Quilis2000} and harassment leads to very little \emph{disc} mass loss for these massive systems \citep{Moore1999}.

\citet{AragonSalamanca2006} were the first to provide a quantitative comparison between S0s and Sps based on the properties of their GCSs. They find that $S_{N}$ is on average a factor $\approx$\,3 larger for the former population than for the latter, and suggest that, under the premise that the number of GCs does not vary, it can be interpreted as evidence for luminosity fading due to star formation shutoff in S0s. The constant $N_{gc}$ hypothesis between the two populations is however questionable, as at least 2/8 of the S0s in their table\,1 have \emph{more} GCs than \emph{any} of the Sps -- and the discrepancy is even larger if Sombrero (NGC\,4594) is not considered to be a spiral (e.g., \citealt{GadottiSJ2012}; \citealt{Spitler2008}; \citealt{Rhode2004}). 
The fading interpretation might thus not be that straightforward for the most massive S0s harbouring populous GCSs -- but is in principle consistent with the origin of less luminous lenticulars. This further supports the existence of two different formation mechanisms for \emph{central} and \emph{satellite} S0s, as recently proposed by \citet{Wilman2012}.

While a detailed comparison between the GCS properties of S0s and Sps is beyond the scope of this paper, we note that Fig.\,\ref{fig:harass_lsb} also shows a slight hint of S0s having a higher \emph{mass} specific frequency than late-type disc galaxies. Uncertainties in both stellar masses and the number of GCs are however significant compared to the differences, and a more detailed study complemented with independent observables (e.g., bulge-to-disc mass fractions, stellar population ages and metallicities) is highly desirable in order to clarify the origin of S0s.

\label{lastpage}

\end{document}